\documentclass{ws-ijmpe}

\begin{document}

\markboth{M. V. T. Machado}{Geometric scaling in ultrahigh energy neutrinos and nonlinear perturbative QCD}

\catchline{}{}{}{}{}

\title{GEOMETRIC SCALING IN ULTRAHIGH ENERGY NEUTRINOS AND NONLINEAR PERTURBATIVE QCD}

\author{\footnotesize M. V. T. MACHADO}

\address{Instituto de Fisica, Universidade Federal do Rio Grande do Sul\\
Av. Bento Gon\c{c}alves 9500, CEP 91501-970. Porto Alegre, RS, Brazil\\
magnus@if.ufrgs.br}



\maketitle

\begin{history}
\received{(received date)}
\revised{(revised date)}
\end{history}

\begin{abstract}
The ultrahigh energy neutrino cross section is a crucial ingredient in the calculation of the event rate in high energy neutrino telescopes. Currently there are several approaches which predict different behaviors for its magnitude for ultrahigh energies. In this contribution is presented a summary of current predictions based on the non-linear QCD evolution equations, the so-called perturbative saturation physics. In particular, predictions are shown based on the parton saturation approaches and the consequences of  geometric scaling property at high energies are discussed. The scaling property allows an analytical computation of the neutrino scattering on nucleon/nucleus at high energies, providing a theoretical parameterization.
\end{abstract}

\section{Introduction}

The investigation of ultrahigh energy (UHE) cosmic neutrinos provides an opportunity for study particle physics beyond the reach of the LHC \cite{neu_review}. As an example, nowadays the Pierre Auger Observatory  is sensitive to neutrinos of energy $\ge 10^8$ GeV \cite{pao}. A crucial ingredient in the calculation of attenuation of neutrinos traversing the Earth and the event rate in high energy neutrino telescopes
is the  high energy neutrino-nucleon cross section, which  provides a probe of Quantum Chromodynamics (QCD) in the kinematic region of very small values of Bjorken-$x$.
The typical $x$ value probed is $x \approx m_W^2/2m_NE_{\nu}$, which implies that for $E_{\nu} \approx 10^8 - 10^{10}$ GeV one have $x \approx 10^{-4} - 10^{-6}$ at $Q^2 \approx 10^4$ GeV$^2$. This kinematical range was not explored by the HERA measurements of the structure functions \cite{hera}.

The description of QCD dynamics in such very high energy limit is a subject of intense debate \cite{hdqcd}. Theoretically, at high energies (small Bjorken-$x$)  one
expects the transition of the regime described by the linear
dynamics, where only the parton emissions are considered, to a new
regime where the physical process of recombination of partons becomes
important in the parton cascade and the evolution is given by a
non-linear evolution equation.  This regime is characterized by the
limitation on the maximum phase-space parton density that can be
reached in the hadron wavefunction (parton saturation), with the
transition being specified  by a typical scale, which is energy
dependent and is called saturation scale $Q_{\mathrm{s}}$.
Moreover,  the growth of the parton distribution is expected to saturate, forming a  Color Glass Condensate (CGC), whose evolution with energy is described by an infinite hierarchy of coupled equations for the correlators of  Wilson lines \cite{hdqcd}.
In the mean field approximation, the first equation of this  hierarchy decouples and boils down to a single non-linear integro-differential  equation: the Balitsky-Kovchegov (BK) equation \cite{BAL,KOVCHEGOV}. Experimentally, possible signals of parton saturation have already
been observed both in  $ep$ deep inelastic scattering at HERA and in deuteron-gold
collisions at RHIC \cite{hdqcd}.

Currently, there are predictions of the neutrino nucleon cross sections with structure functions constrained by HERA data are based on linear  dynamics \cite{ccs}, using DGLAP or an unified DGLAP/BFKL evolution, or phenomenological models that resembles the expected behavior predicted by  the non-linear QCD dynamics \cite{bhm} (i.e., the proton structure function saturating the Froissart bound at asymptotic energies, $F_2^p \propto \ln^2 (1/x)$). As a general feature, the nonlinear QCD dynamics predicts sizable suppression of UHE neutrino cross section in comparison with standard approaches. Here, we summarize the main results of works presented in Refs. \cite{magno1,magno2}. In Ref. \cite{magno1} the contribution of non-linear effects  was estimated considering distinct phenomenological models based on saturation physics. An update on those calculations have been done recently \cite{victor}. In Ref. \cite{magno2}, the geometric scaling property (which is a natural consequence of the asymptotic solutions of the nonlinear QCD evolution equations) is considered to obtain an analytical parameterization for the UHE neutrino cross sections. In what follows we introduce the theoretical and phenomenological tools and present the main results and predictions

\section{The UHE neutrino cross section and nonlinear perturbative QCD approaches}
\label{cross}

Deep inelastic neutrino scattering is described in terms of charged current (CC) and neutral current (NC) interactions, which proceed through $W^{\pm}$ and $Z^0$  exchanges, respectively. The total cross sections are given by:
\begin{eqnarray}
& & \sigma_{\nu N}^{CC,\,NC} (E_\nu)  = \int_{Q^2_{min}}^s dQ^2 \int_{Q^2/s}^{1}  \frac{dx}{x s}\,
\frac{G_F^2 M E_{\nu}}{\pi} \left(\frac{M_i^2}{M_i^2 + Q^2}\right)^2 \nonumber \\
&\times & \left[\frac{1+(1-y)^2}{2} \, F_2^{CC,\,NC} - \frac{y^2}{2}F_L^{CC,\,NC}  + y (1-\frac{y}{2})xF_3^{CC,\,NC}\right],
\label{difcross}
\label{total}
\end{eqnarray}
where $E_{\nu}$ is the neutrino energy, $s = 2 ME_{\nu}$ with $M$ the nucleon mass, $y = Q^2/(xs)$ and $Q^2_{min}$ is the minimum value of $Q^2$ which is introduced in order to stay in the deep inelastic region. In what follows we assume $Q^2_{min} = 1$ GeV$^2$. Moreover,
$G_F$ is the Fermi constant and $M_i$ denotes the mass of the charged of neutral gauge boson.

In the QCD improved parton model the structure functions $F_i(x,Q^2)$ are expressed in terms of the parton distributions on the nucleon, which satisfy the DGLAP  \cite{dglap} linear dynamics. On the other hand, an efficient way of introducing non-linear effects is the description of the structure functions considering the color dipole approach in which the DIS to low $x$ can be viewed as a result of the interaction of a color $q\bar{q}$ dipole which the gauge boson fluctuates \cite{nik}.  In this approach the $F_2^{CC,\,NC}$ structure function is expressed in terms of the transverse and longitudinal structure functions, $F_2^{CC,\,NC}=F_T^{CC,\,NC} + F_L^{CC,\,NC}$ which are given by
\begin{eqnarray}
F_{T,L}^{CC,\,NC}(x,Q^2)  =  \frac{Q^2}{4\pi^2} \int_0^1 dz \int d^2  r |\Psi^{W,Z}_{T,L}(r,z,Q^2)|^2 \sigma^{dip}(r,x)\,\,
\label{funcs}
\end{eqnarray}
where $r$ denotes the transverse size of the dipole, $z$ is the longitudinal momentum fraction carried by a quark and  $\Psi^{W,Z}_{T,L}$ are proportional to the wave functions of the virtual charged of neutral gauge bosons corresponding to their transverse or longitudinal polarizations. Explicit expressions for $\Psi^{W,Z}_{T,L}$ are given, e.g., in Ref. \cite{magno1}.    Furthermore, $\sigma^{dip}$ describes the interaction of the  color dipole with the target. In next section we will discuss some models for $\sigma^{dip}$, based on the non-linear QCD dynamics, which describe the current HERA data. As a comment, the DESY-HERA measurements of the structure function  at low - $x$ ($x  \approx 10^{-6}$) are for very low values of $Q^2$ ($Q^2 \ll 1$ GeV$^2$), which implies that  small $x$ extrapolations of the parton distributions are necessary to estimate $\sigma_{\nu N}$ above $E_{\nu} \approx 10^7$ GeV.

As mentioned above, UHE neutrino-nucleon cross section accesses very large values of $Q^2$ and very small values of Bjorken $x$. The usual linear perturbative QCD evolution equations  predictions are based on the linear DGLAP and/or BFKL equations which implies a power increase with the energy  of the neutrino-nucleon cross section that eventually violate the Froissart bound. A easy way to introduce consistently the unitarity constraints on the theoretical estimate of the ultrahigh energy behavior of the neutrino-nucleon cross section is to express the structure functions in the dipole approach [Eq. (\ref{funcs})] and to consider the state-of-art of the non-linear QCD dynamics: the Color Glass Condensate formalism. In this formalism, the dipole - target cross section $\sigma^{dip}$ can be computed in the eikonal approximation, resulting
\begin{eqnarray}
\sigma^{dip} (x,r)=2 \int d^2 b \, {\cal{N}}(x,r,b)\,\,,
\end{eqnarray}
where ${\cal{N}}$ is the  dipole-target forward scattering amplitude
for a given impact parameter $b$  which encodes all the
information about the hadronic scattering, and thus about the
non-linear and quantum effects in the hadron wave function. It is
useful to assume that the impact parameter dependence of $\cal{N}$
can be factorized as ${\cal{N}}(x,r,b) = {\cal{N}}(x,r)
S(b)$, so that $\sigma^{dip}(x,r) = {\sigma_0}
\,{\cal{N}}(x,r)$, with $\sigma_0$ being   a free parameter
related to the non-perturbative QCD physics. The Balitsky-JIMWLK
hierarchy  describes the energy evolution of the dipole-target
scattering amplitude ${\cal{N}}(x,r)$.
In the mean field approximation, the first equation of this  hierarchy decouples and boils down to the Balitsky-Kovchegov (BK) equation \cite{BAL,KOVCHEGOV}. The dipole-target cross section  can also be calculated considering  phenomenological parameterizations for ${\cal{N}}(x,r)$ based on saturation physics, which provide
an economical description of a wide range of data with a few parameters. This approach has been considered in Refs. \cite{magno1,victor}. In general, the   dipole scattering amplitude is modeled in the coordinate space in terms of a simple Glauber-like formula as follows
\begin{eqnarray}
{\cal{N}}(x,r) = 1 - \exp\left[ -\frac{1}{4} (r^2 Q_s^2)^{\gamma (x,r^2)} \right] \,\,,
\label{ngeral}
\end{eqnarray}
where   $\gamma$ is the anomalous dimension of the target gluon distribution.    The main difference among the distinct phenomenological models comes from the  predicted behavior for the anomalous dimension, which determines  the  transition from the non-linear to the extended geometric scaling regimes, as well as from the extended geometric scaling to the DGLAP regime \cite{hdqcd}.

One important property of the  nonlinear perturbative QCD  approaches for high energy deep inelastic $ep(A)$ scattering   is the prediction of the  geometric scaling. Namely,  the total $\gamma^* p(A)$ cross section at large energies is not a function of
the two independent variables $x$ and $Q^2$, but is rather a
function of the single variable $\tau_A = Q^2/Q_{\mathrm{sat,A}}^2$. As usual, $Q^2$ is the photon virtuality and $x$ the Bjorken variable. Geometric scaling is the exact asymptotic solution of a general class of nonlinear evolution equations and it appears as a universal property of these kind of equations.  The saturation momentum $Q_{\mathrm{sat,A}}^2(x;\,A)\propto\frac{xG_A(x,\, Q_{\mathrm{sat}}^2)}{\pi R_A^2} \simeq A^{\,\alpha}\,x^{-\lambda}$ ($\alpha \simeq 1/3$, $\lambda \simeq 0.3$)  is connected with the phenomenon of gluon saturation. In principle, geometric scaling is predicted to be present only on process dominated by low momenta. However, it is known that the geometric scaling is preserved by the QCD evolution  up
to relatively large virtualities \cite{hdqcd}, within the kinematical window $Q_{\mathrm{sat}}^2 (x) < Q^2 < Q_{\mathrm{sat}}^4 (x)/\Lambda^2_{\mathrm{QCD}}$. That is, the scaling property extends towards very large virtualities provided one stays in low-$x$. This kinematical window is further enlarged due to the nuclear enhancement of the saturation scale. These facts have direct consequences in the behavior of UHE neutrino cross section.

As similar scaling holds for the massive boson-nucleon cross section, we are able to compute analytically the neutrino cross sections. This feature was verified to be true in Ref. \cite{mairon}, where geometric scaling was shown to occur in the small-$x$ charged current neutrino data. In this sense, we are able to construct a theoretical prediction which is model independent. They are as follows \cite{magno2},
\begin{eqnarray}
\sigma^{\mathrm{CC,\,NC}}_{(\nu,\,\bar{\nu})}  =  {\cal N}_{(i)} \,A^{\alpha}\left(\frac{R_A^2}{R_p^2}\right)^{1-\alpha}\!\left[ C_1^{(i)}\,E_{\nu}^{\,\omega_{\mathrm{scal}}} - C_2^{(i)}\right]\,,
\label{resgs}
\end{eqnarray}
where ${\cal N}_{(i)}$ are overall normalizations, $C_{1,\,2}^ {(i)}$ are  numerical constants with $i=\mathrm{CC,\,NC}$, $\omega_{\mathrm{scal}}=b\lambda$ and $\alpha = b/\delta$. This implies in a mild power-like rise $\omega_{\mathrm{scal}}\simeq 0.2$ for the neutrino cross section in contrast with other theoretical approaches. The nuclear dependence is approximately linear, $\sigma_{\nu,\bar{\nu}}^{\mathrm{nuclei}}\propto A\,\sigma_{\nu,\bar{\nu}}^{\mathrm{nucleon}}$,  once $b\simeq \delta$ and hence $\alpha \approx 1$.  The remaining constants are given by (see Ref. \cite{magno2} for details),
\begin{eqnarray}
& &  {\cal N}_{(i)}   =  R_{\mathrm{cor}}\left(\frac{\bar{\sigma}_0\,G_F^2\,M_{W,Z}^2}{8\pi^3\lambda}\right)\left(\frac{a\,x_0^{\,\omega_{\mathrm{scal}}}}{b}\right)\frac{B_{(i)}}{\alpha_{\mathrm{em}}\,\sum_f e_f^2}\,,\\
& & C_1^{(i)}  = \pi\, (2\,m_N)^{\omega_{\mathrm{scal}}}\left(M_{W,Z}^2\right)^{-\nu_{\mathrm{scal}}}\mathrm{csc}\left(\pi \,\nu_{\mathrm{scal}}\right)\,\left(1-\nu_{\mathrm{scal}} \right),\,C_2^{(i)} = \pi\, \left(M_{W,Z}^2\right)^{-b}\mathrm{csc}\left(\pi \,b\right)\,\left(1-b \right)\nonumber ,
\end{eqnarray}
where one uses the notation $B_{\mathrm{CC}} = 4$, $B_{\mathrm{NC}} = K_{\mathrm{chiral}}$ and $\nu_{\mathrm{scal}}=b-\omega_{\mathrm{scal}}$. Numerically, this gives a total cross section $\sigma_{\nu,\bar{\nu}}^{tot}= 1.48\times10^{-34}\,A^{\alpha}(E_{\nu}/\mathrm{GeV})^{0.227}$ cm$^2$. The cross section above can have implications for neutrino observatories because experiments are planned to detect UHE by observation of the nearly horizontal air showers in Earth coming from neutrino-air interactions \cite{ancho2}. A reduced cross section produces a smaller event rate for such neutrino-induced showers and could compromise the detection signal. However, the rate of up-going air showers initiated by muon and tau leptons produced in neutrino-nucleon reactions just below the surface would increase, being possibly larger than the horizontal air shower rate.

\section{Results and summary}
\label{resultados}

\begin{figure}[t]
\centerline{\psfig{file=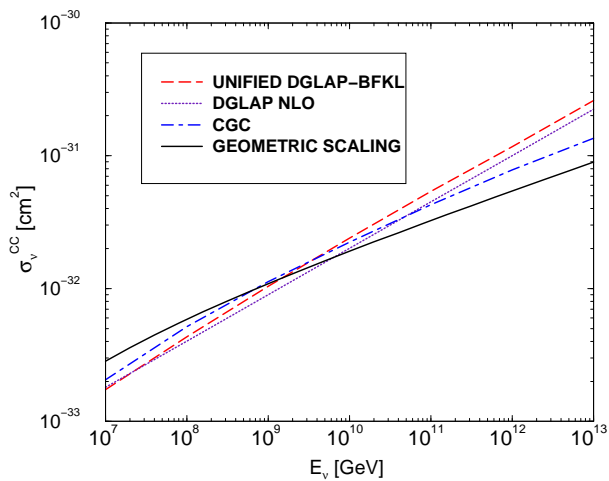,width=8cm}}
\vspace*{8pt}
\caption{Comparison among perturbative QCD approaches: unified DGLAP-BFKL, NLO DGLAP and a phenomenological model based on the Color Glass Condensate physics (CGC). The analytical geometric scaling parameterization is also presented.}
\label{fig:1}
\end{figure}

Let us present the comparison between the predictions of the linear approaches (NLO DGLAP and unified DGLAP/BFKL) and non-linear QCD approaches (a phenomenological model for the Color Glass Condensate and the geometric scaling prediction). In  Fig. \ref{fig:1} the  energy dependence of the neutrino nucleon CC cross section predicted by the  linear and nonlinear perturbative QCD approaches are compared.  As expected from the solution of the DGLAP equation at small-$x$, the NLO DGLAP (dot-dashed curve) and unified DGLAP/BFKL (long-dashed curve) results predict a strong increase of the cross section at ultrahigh energies.  Such estimates are largely used  to estimate the event rates in neutrinos telescopes. Clearly, the CGC calculation predicts an increasingly suppressed  cross section compared to its linear QCD counterpart.

In Fig. \ref{fig:1} the solid line represents the result of geometric scaling parameterization, Eq. (\ref{resgs}).  The suppression in the cross section is even higher than for the CGC phenomenological models reaching a factor two compared the the linear QCD approaches. The present results demonstrate that the determination of $\sigma_{\nu N}$ can be useful to constrain the underlying QCD dynamics. In principle, this cross section could be  constrained at high energies by studying the  ratio between quasi-horizontal deeply penetrating air showers and Earth-skimming tau showers \cite{ancho2}.

As a summary, the detection of UHE neutrinos may shed light on the observation of air showers events with energies in excess of $10^{11}$ GeV, reveal aspects of new physics as well as of the QCD dynamics at high energies. One of the main ingredients for estimating event rates in neutrino telescopes and  cosmic ray observatories (e.g. AUGER)
is the neutrino - nucleon cross section. We have examined to what extent the cross section is sensitive to the presence of new dynamical effects in the QCD evolution.
We compare the predictions of several approaches based on different assumptions for the QCD dynamics. In particular, we have compared the phenomenological models based on nonlinear perturbative QCD  with the usual NLO DGLAP evolution. Moreover, we compute a model independent cross section, using the experimental verification of the geometric scaling property.  The results demonstrate that the current theoretical uncertainty for the neutrino-nucleon cross section reaches a factor two or higher for neutrino energies above  $10^{11}$ GeV.

\end{document}